\documentclass[aps,prb,twocolumn,showpacs,10pt,superscriptaddress,floatfix,longbibliography]{revtex4-1}

\usepackage{newtxtext,newtxmath}
\usepackage{graphicx}
\usepackage{subfigure}
\usepackage{amsmath}
\usepackage{amsfonts}
\usepackage{amssymb}
\usepackage{mathrsfs}
\usepackage{bbm}
\usepackage{subfigure}
\usepackage{xcolor}
\usepackage[colorlinks=true]{hyperref}
\usepackage{ulem}
\usepackage[T1]{fontenc}

\begin{document}
\title{Josephson radiation from nonlinear dynamics of Majorana zero modes}

\author{Jia-Jin Feng}
\altaffiliation{These authors contributed equally to this work.}
\affiliation{School of Physics, Sun Yat-sen University, Guangzhou 510275, China}
\affiliation{International Center for Quantum Materials, Peking University, Beijing 100871, China}

\author{Zhao Huang}
\altaffiliation{These authors contributed equally to this work.}
\affiliation{Theoretical Division, Los Alamos National Laboratory, Los Alamos, New Mexico 87545, USA}
\affiliation{Texas Center for Superconductivity, University of Houston, Houston, Texas 77204, USA}

\author{Zhi Wang}
\email{wangzh356@mail.sysu.edu.cn}
\affiliation{School of Physics, Sun Yat-sen University, Guangzhou 510275, China}

\author {Qian Niu}
\affiliation{Department of Physics, The University of Texas at Austin, Austin, Texas 78712, USA}

\begin{abstract}
Josephson radiation is a powerful method to probe Majorana zero modes in topological superconductors. Recently, Josephson radiation with half the Josephson frequency has been experimentally observed in a HgTe-based junction, possibly from Majorana zero modes. However, this radiation vanishes above a critical voltage, sharply contradicting previous theoretical results. In this work, we theoretically obtain a radiation spectrum quantitatively in agreement with the experiment after including the nonlinear dynamics of the Majorana states into the standard resistively shunted junction model. We further predict two new structures of the radiation spectrum for future experimental verification: an interrupted emission line and a chaotic regime.
We develop a fixed-point analysis to understand all these features. Our results 
resolve an apparent discrepancy between theory and experiments, and will inspire reexamination of structures in radiation spectra of various topological Josephson junctions.
\end{abstract}
\date{\today}
\maketitle

\section{Introduction}
The fault-tolerant quantum storage and operation is one of the promising schemes for achieving quantum computation\cite{kitaev2003anyon}. It is based on the premises of non-Abelian statistics, which is associated with the Majorana zero modes in the topological superconductors \cite{alicea2012new}. Identifying the Majorana zero modes in realistic experimental setups remains a challenge\cite{lutchyn2018review}. Among various methods,\cite{beenakker2013,aguado2017,Sato2017}, the Josephson effect has attracted considerable attentions due to its phase sensitive nature\cite{tanaka2009,ioselevich2011,oostinga2013,matthews2014,sothmann2016,peng2016TJJ,deacon2017radiation,cayao2017,kamata2018,schrade2018,laroche2019radiation,ren2019josephson,stern2019,trimble2019}.
The Majorana zero modes in a Josephson junction carry a nontrivial Josephson current with $4\pi$-periodicity in Josephson phase \cite{kitaev2001unpaired,Kwon2004}, which is in contrast to the $2\pi$-periodic Josephson current in conventional junctions.
However, its direct measurement is hindered by the coupling to other distant Majorana zero modes or other quasiparticle excitations\cite{fu2009QSHIjunction,badiane2011,houzet2013,crepin2014}. The hybridization opens a small gap in the $4\pi$-periodic Andreev levels and breaks the local parity conservation\cite{fu2009QSHIjunction}, reducing the $4\pi$-periodicity to the $2\pi$-periodicity.

\begin{figure}[t]
\begin{center}
\includegraphics[clip = true, width =\columnwidth]{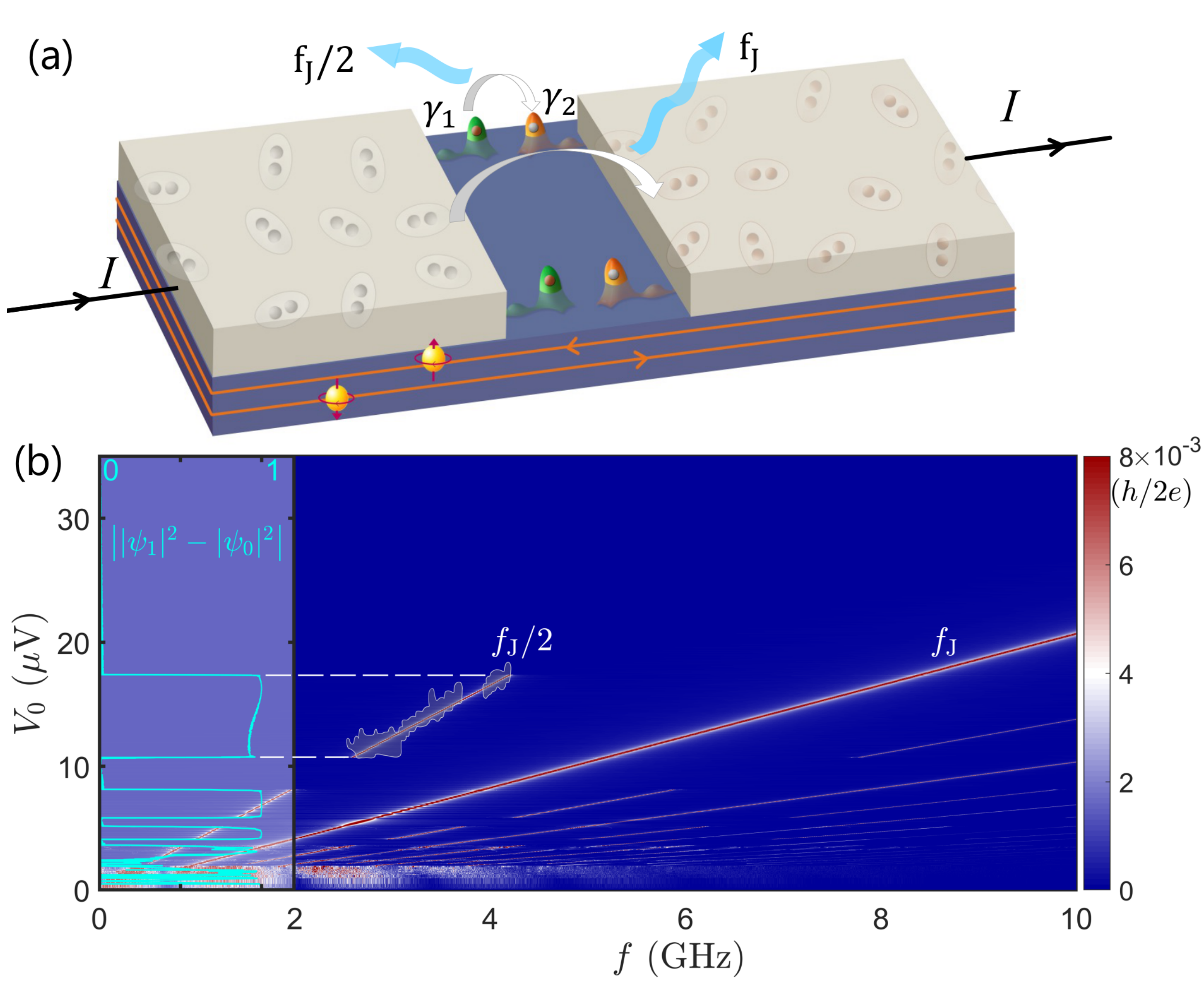}
\caption{ (a) Schematic setup for Josephson radiation from overdamped junction with Majorana zero modes. The junction is formed by two ordinary s-wave superconductors (gray) on top of a quantum-spin-Hall-insulator (blue), whose edge states host four Majorana zero modes. Cooper-pair tunneling emits Josephson radiation with a quantized frequency $f_{\rm J}= 2eV/h$, while the single-electron tunneling through Majorana zero modes emits radiation with the half frequency $f_{\rm J}/2$. (b) Radiation spectra from the numerical simulations of quantum resistively shunted junction model. The $f_{\rm J}/2$ emission line vanishes above a critical voltage, in agreement with recent experimental results (gray patch). Appearance of the $f_{\rm J}/2$ line corresponds to nonzero expectation value $|\psi_1|^2-|\psi_0|^2$ of the two-level system from Majorana zero modes, as illustrated with the green line at the left. Parameters are $E_{\rm M}=19.6u$eV, $\delta=0.196u$eV, $R=50\Omega$, $I_{\rm M} =16$nA and $I_{\rm J}=63$nA in Eqs.~(\ref{eq:linearSE}) and (\ref{eq:RSJ}), which are realistic values close to the estimations in the experiment\cite{deacon2017radiation}. We scan $I_{\rm in}\in$ [0, 0.7$\mu$A] to obtain radiation spectra for $V_0\in$ [0, 35$\mu$V].}
\label{fig:setup}
\end{center}
\end{figure}

To circumvent this difficulty one can drive nonequilibrium dynamics in topological Josephson junctions\cite{pikulin2012,sanjose2012,rokhinson2012fractional,dominguez2012,wiedenmann2016shapiro4pi,bocquillon2017gapless,dominguez2017,pico2017,sun2018,murani2019}, extracting $4\pi$-periodic information from the electromagnetic radiation of the junction\cite{deacon2017radiation,kamata2018,laroche2019radiation}. An extra emission line in the spectrum function with half of the Josephson frequency $f_{\rm J}/2$ was expected as a result of the $4\pi$-periodic Josephson relation from Majorana zero modes\cite{dominguez2017}, as illustrated in Fig.~\ref{fig:setup}a. In a recent experiment Ref. [\onlinecite{deacon2017radiation}], this emission line was indeed observed in a HgTe-based topological junction. However, it mysteriously vanishes above a critical voltage, which is inconsistent with previous theories on the Josephson radiation\cite{deacon2017radiation}. This discrepancy endangers the claim of the Majorana zero modes in such topological junctions.

In this work, we reveal that the Josephson radiation spectra found in the experiments can be well explained after considering a rarely noticed correlation between the nonlinear dynamics of the Josephson phase and the time evolution of Majorana states. By including the Majorana states into the standard resistively shunted junction model, we show that this correlation induces an exact cancellation of the $4\pi$-periodic Josephson current for voltages above a critical value, which leads to the vanishment of the $f_{\rm J}/2$ radiation. In particular, our results show a quantitative agreement with the experimental data. For a better understanding, we cast the model into an equivalent classical model with three nonlinear equations, and use the method of averaging to obtain the fixed-point portrait. We find that the vanishment of the emission line can be well characterized by behaviors of the fixed points. We also predict new interrupted emission lines and chaotic dynamics in parameter regimes not yet experimentally explored. Our results highlight the rich physics stemming from the interplay between nonlinear dynamics and nontrivial topology in quantum materials.

\section{Superposition of Majorana states with opposite parities}
Before theoretically demonstrating the vanishment of the emission line with the frequency $f_{\rm J}/2$, we first come to the conventional wisdom for the analysis of Josephson radiation from the $4\pi$-periodic Josephson effect, and see its limitations in explaining existing experiments.

For a topological Josephson junction, it is well known that the Majorana zero modes carry the Josephson current\cite{kitaev2001unpaired,fu2009QSHIjunction} with a $4\pi$-periodic current-phase relation $I \propto \pm \sin \theta/2$, where $\theta$ is the Josephson phase and the plus/minus sign represents the fermionic parity of coupled two Majorana zero modes. If we consider a dc voltage $V_0$, the Josephson phase increases linearly with time according to the ac Josephson relation $\theta (t) =2 eV_0 t/\hbar$, and leads to $I\propto\pm \sin eV_0t/\hbar$, which induces electromagnetic radiation with the frequency of $eV_0/h$. It is exactly half of the conventional quantized Josephson frequency $f_{\rm J}=2eV_0/h$.
Pictures may become even clearer from a quantum point of view. The radiation with frequency $f_{\rm J}$ represents the energy loss for a Cooper pair to tunnel through a junction biased with voltage $V_0$, while the radiation with frequency $f_{\rm J}/2$ corresponds to the coherent single-electron tunneling through the Majorana channel, as illustrated in Fig.~\ref{fig:setup}a. It is important to see whichever picture we take, the $f_{\rm J}/2$ radiation is predicted to exist for all voltages by the theory, which thus cannot explain its vanishment above a critical voltage in the recent experiment.

The central idea of solving this discrepancy lies at the plus/minus sign in front of the current-phase relation. At first glance, one can take either sign since the radiation spectra are identical. A more careful examination, however, brings to light a key observation that the sign itself can be time-dependent, and correlate with the dynamics of the Josephson phase, possibly causing a dramatic modification to the radiation spectra.
To correctly take this phenomenon into consideration, the fermionic parity of the Majorana zero modes must be examined in more detail. The two Majorana zero modes $\gamma_1$ and $\gamma_2$ define a parity operator $\hat s_z = i\gamma_1 \gamma_2$ which has two eigenstates, $\hat s_z |0\rangle = -|0\rangle$ and $\hat s_z |1\rangle = |1\rangle$. In the junction these two states constitute a typical two-level system and the system can stay at the superposition state $|\psi \rangle = \psi_0 |0\rangle + \psi_1 |1\rangle$. The supercurrent through the Majorana zero modes should be determined by the expectation value of the parity operator\cite{fu2009QSHIjunction,feng2018hysteresis}
\begin{eqnarray}
 I= I_{\rm M} \langle \psi| \hat s_z |\psi \rangle \sin ({eV_0t}/ {\hbar})
 \end{eqnarray}
 with $\langle \psi| \hat s_z |\psi \rangle=|\psi_1|^2 - |\psi_0|^2$.
 
Now the physic is clear. If $|\psi\rangle$ stays on the eigenstate $|0\rangle$ or $|1\rangle$, we have $\langle \psi| \hat s_z |\psi \rangle=\pm 1$ and the $4\pi$-periodic emission line exists for all voltages, which contradicts the experimental results. However, if $|\psi\rangle$ is a superposition state with $|\psi_0|=|\psi_1|$, we have $\langle \psi| \hat s_z |\psi \rangle=0$ which means the zero current through the Majorana channel. This naturally explains the experimentally observed vanishment of the $4\pi$-periodic radiation.

\section{Numerical simulation of experimental observation}
 The complete model for the dynamics in this junction requires the inclusion of a dynamical equation for the two-level system. This has been established in a quantum resistively shunted junction model\cite{huang2015PRA,feng2018hysteresis}. For a HgTe-based junction, a minimal model requires two ordinary s-wave superconductors on top of a quantum-spin-Hall-insulator\cite{fu2009QSHIjunction}, as shown in Fig.~\ref{fig:setup}a. 
When we consider the total parity conservation\cite{supplemental,feng2018hysteresis}, we can obtain the Schr\"{o}dinger equation for the two-level system as,
\begin{eqnarray}\label{eq:linearSE}
i \hbar \frac{{\rm d}}{{\rm d} t} \begin{pmatrix}
\psi_0  \\
\psi_1
\end{pmatrix} = \begin{pmatrix}
E_{\rm M}\cos{\frac{\theta}{2} } & \delta\\
\delta & -E_{\rm M}\cos{\frac{\theta}{2} }
\end{pmatrix} \begin{pmatrix}
\psi_0  \\
\psi_1
\end{pmatrix},
\end{eqnarray}
where $E_{\rm M}$ represents the Josephson energy from the coupling between the Majorana zero modes at the left and right interface, and $\delta$ is the hybridization energy from the wave function overlap between the Majorana zero modes at the upper and lower edge in Fig.~\ref{fig:setup}a\cite{supplemental}. 
We notice that the Josephson phase $\theta$ is a variable of the Hamiltonian of the two-level system, meaning that the Josephson phase dynamics directly influence the evolution of the wave function.

By solving the Schr\"{o}dinger equation, we can obtain the Josephson current carried by the Majorana zero modes. Adding this tunneling current to the resistively shunted junction equation\cite{tinkham2004}, we arrive at the dynamical equation for the Josephson phase,
\begin{eqnarray}\label{eq:RSJ}
\dot \theta = \frac{2eR}{\hbar} \left[I _{\rm in}- I_{\rm J} \sin \theta - I_{\rm M} (|\psi_1|^2 - |\psi_0|^2) \sin \frac {\theta}{2}\right],
\end{eqnarray}
where $R$ is the resistance of the junction, $I_{\rm in}$ is the injected current, $I_{\rm J}$ represents the maximum $2\pi$-periodic supercurrent from the Cooper-pair tunneling\cite{mahan2000}, $I_{\rm M}$ represents the maximum $4\pi$-periodic supercurrent from the single charge tunneling through Majorana zero modes.
The two equations (\ref{eq:linearSE}) and (\ref{eq:RSJ}) couple the Josephson phase $\theta$ and the Majorana state $|\psi \rangle$, and must be solved simultaneously to obtain the full dynamics of the junction.
This is a minimal model for studying the dynamics of the topological Josephson junctions. It can be interpreted as equations of motion for a particle with a pseudo spin one-half, subject to a one-dimensional spin-dependent potential. It has been used to successfully explain the nontrivial hysteretic I-V curves in topological Josephson junctions\cite{oostinga2013,wiedenmann2016shapiro4pi,feng2018hysteresis}.

Analysis of radiation spectra is based on the solution of Eq.~(\ref{eq:linearSE}) and (\ref{eq:RSJ}).
Upon the injected current $I_{\rm in}$ larger than the critical current, the Josephson phase begins to increase with an oscillating velocity, which induces both dc and ac voltage by checking $V(t) = \hbar \dot \theta(t)/2e$. The dc voltage pumps in energy and the ac voltage emits out electromagnetic energy. Numerically, we take Fourier transformation to obtain the spectrum function $v(f,I_{\rm in})=\int V(t,I_{\rm in})e^{-i2\pi ft} {\rm d}t$ by sweeping the values of $f$ and $I_{\rm in}$. Here the zero-frequency spectrum function is the dc voltage $V_0 \equiv v(f=0,I_{\rm in})$. We then rewrite the finite frequency spectrum as a function of frequency and dc voltage $v(f,V_0)$ to compare with experiments. When $|\psi_1|^2 - |\psi_0|^2$ is fixed and nonzero, we can analytically obtain $V(t)=\sum_n v(nf_{\rm J}/2)e^{i2\pi nf_{\rm J}/2}$ which has quantized frequencies \cite{supplemental}.

We have specifically chosen realistic parameters to obtain the emission lines shown in Fig.~\ref{fig:setup}b. Several straight emission lines are presented, corresponding to the quantized radiation frequencies $f_{\rm J}/2$, $f_{\rm J}$, and higher harmonics.
All the emission lines are sharp and straight, consistent with the quantization feature of the Josephson radiation appeared in conventional junctions.
The surprise comes from the $f_{\rm J}/2$ emission line. Our simulation shows a unique feature: the emission line vanishes above a critical voltage.

This unusual feature is highly relevant to recent experimental observations and beyond the limitations of conventional RSJ model. As reported in a HgTe-based topological junction in Ref. [\onlinecite{deacon2017radiation}], a clear termination of the $f_{\rm J}/2$ emission line is observed. The gray patch in Fig.~\ref{fig:setup}b is the radiation spectrum observed in the experiment for the $f_{\rm J}/2$ frequency, which is dragged out from Fig.~2f in Ref. [\onlinecite{deacon2017radiation}]. With our realistic parameters, we can repeat this experimental data as shown by the solid line overlapping with the shadow. The correlation between the Josephson phase and the wave function of the Majorana state in the dynamics is crucial for this phenomenon.

\section{Prediction of the interrupted emission line and chaotic dynamics}
Besides the successful theoretical repeat of the vanishment of the $f_{\rm J}/2$ Josephson radiation at high voltages, we further find a novel phenomenon: the emission line is interrupted at low voltages. This is especially clear for $f \in$ [0, 2GHz] in Fig.~\ref{fig:setup}b. We intentionally changed the background color in this range because it is outside the detection area of the previous experiment\cite{deacon2017radiation}. We point out that, in the existing experimental data, there is already some slight discontinuity in the $f_{\rm J}/2$ emission line, which might be relevant to our prediction.
In the meantime, we also notice an interesting chaotic behavior at zero voltage limit, shown as noise-like results at the bottom of Fig.~\ref{fig:setup}b. These chaotic behaviors originate from the nonlinearity in the dynamics. We expect the experimental check in the future on these two predictions.

Now let us explore why the radiation vanishes above the critical voltage and the emission line is interrupted at low voltages. For a nonzero voltage, $|\psi_1|^2-|\psi_0|^2$ evolves to a static value if the dynamics is not chaotic. This value as a function of $V_0$ is shown by the left part of Fig.~\ref{fig:setup}b. For certain regimes of $V_0$, this value becomes zero in this dynamics, which leads to zero tunneling current through the Majorana channel. This is the reason for the vanishment of $f_{\rm J}/2$ radiation. Outside these regimes, currents from opposite parities do not cancel ($|\psi_1|^2-|\psi_0|^2\neq 0$), so we can have the $f_{\rm J}/2$ emission line. Therefore, we get the interrupted emission line. We can see clearly the quantitative matching between the zero expectation value and the vanishment of the $f_{\rm J}/2$ emission line. In particular for the large-voltage limit, $|\psi_1|^2-|\psi_0|^2$ is always zero. We thus have a critical voltage above which the emission line is always absent.

\begin{figure}[t]
\begin{center}
\includegraphics[clip = true, width =\columnwidth]{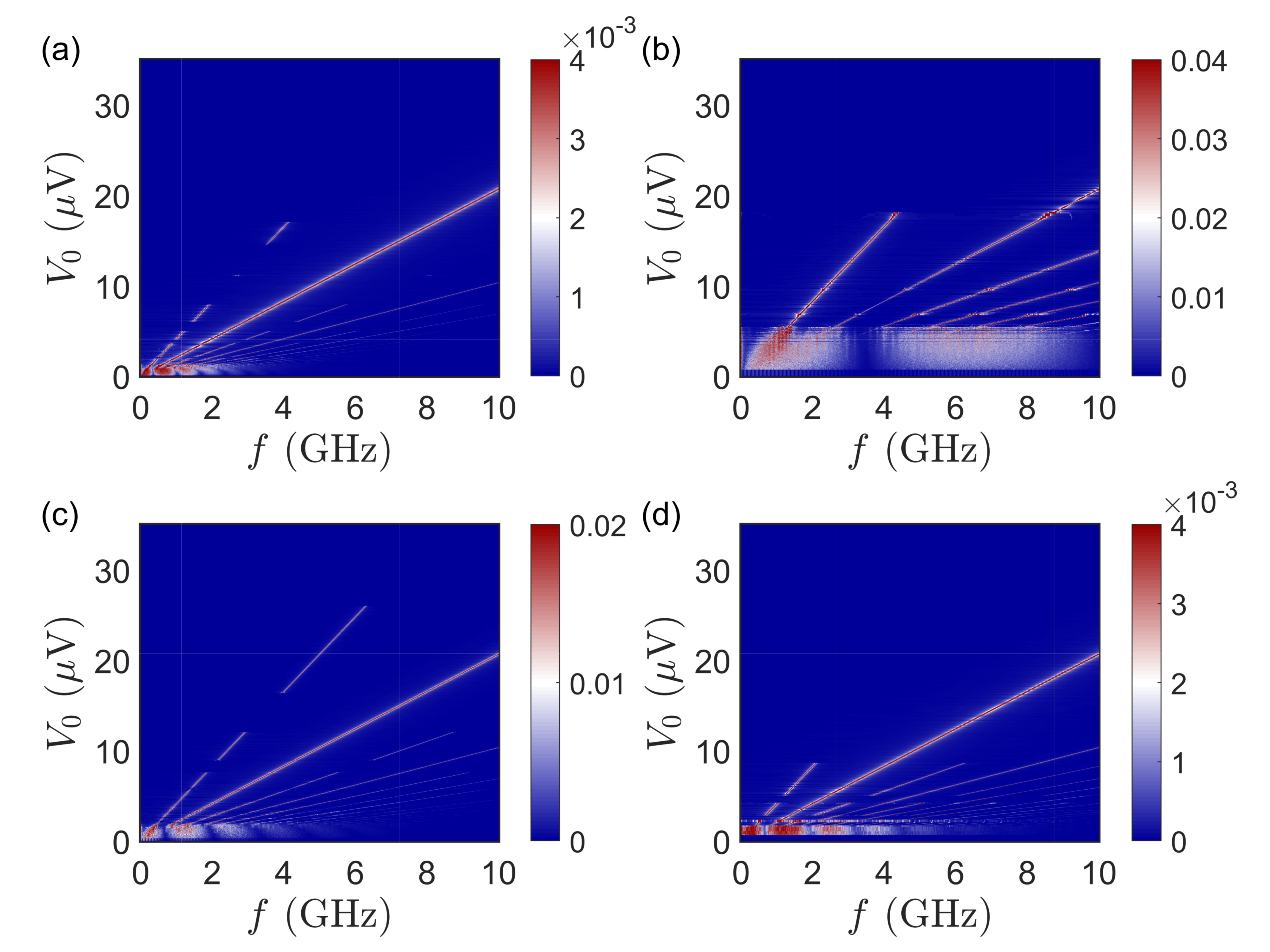}
\caption{Radiation spectra from the numerical simulations of quantum resistively shunted junction model for four typical parameters: (a) $R/R_0 = 0.5$ and (b) $R/R_0 = 5$, (c) $E_{\rm M}/E_{\rm M0} = 1.5$, (d) $E_{\rm M}/E_{\rm M0} = 0.5$, where $R_0$ and $E_{\rm M 0}$ represent the parameters used in Fig.~\ref{fig:setup}b, while all other parameters are taken the same.}
\label{fig:spectra}
\end{center}
\end{figure}

Now we investigate how the above features change with experimentally controllable parameters. In Figs. \ref{fig:spectra}a and \ref{fig:spectra}b we change the resistance to half and ten times of the value used in Fig.~\ref{fig:setup}b.
We find that changing the resistance does not influence the critical voltage. However, the interrupted feature is significantly modulated.
The reduction of the resistance shortens the emission segments, while increasing the resistance results in elongation of the segments. For a large resistance, as shown in Fig.~\ref{fig:spectra}b., the segments fuse into a single emission line. In this limit, we cannot see the interruption any more.
In Figs. \ref{fig:spectra}c and \ref{fig:spectra}d we modulate the Josephson energy $E_{\rm m}$, to half and twice of the value used in Fig.~\ref{fig:setup}b. We find that the critical voltage is proportional to the Josephson energy, which is suitable for experimental check since the Josephson energy can be easily modulated by orders with a gate voltage.
These results for different parameters provide a detailed guidance to check the structure of the Josephson radiations from the $4\pi$-periodic Josephson relation experimentally.

\begin{figure}[t]
\begin{center}
\includegraphics[clip = true, width =\columnwidth]{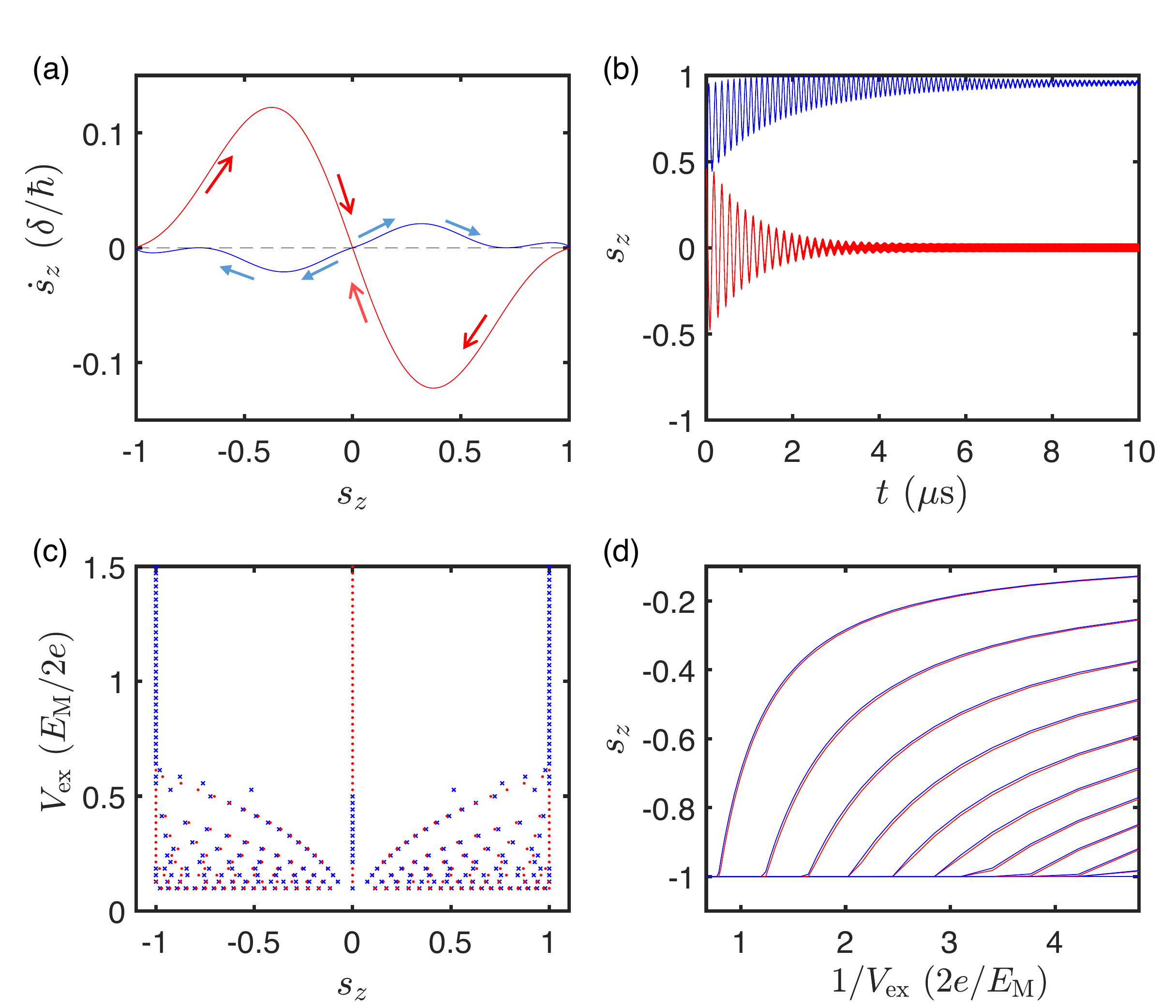}
\caption{(a) Phase-space flow for two typical excess voltages $V_{\rm ex} = 0.556 E_{\rm M}/2e $ (red curve) and $V_{\rm ex} = 0.498 E_{\rm M}/2e$ (blue curve). (b) Time evolution of $s_z=|\psi_1|^2 - |\psi_0|^2$ for the high voltage case $V_0 = 20\mu$eV (red curve) and the low voltage case $V_0 = 14\mu$eV (blue curve).  (c)
Demonstration of stable fixed points (red dots) and unstable fixed points (blue crosses) under the variation of the excess voltages. Other parameters are taken the same as in Fig.~\ref{fig:setup}b.
(d) Illustration of the bifurcation of fixed points with parameters same as Fig.~\ref{fig:spectra}b.}
\label{fig:bifurcation}
\end{center}
\end{figure}

\section{Theoretical understanding based on fixed-point analysis}
The key ingredient to our explanation of the vanishing $f_{\rm J}/2$ emission line is the damping of the expectation value $|\psi_1|^2-|\psi_0|^2$ which causes a perfect cancellation of the $4\pi$-periodic supercurrent. This quantum damping can be understood more transparently by casting the quantum resistively shunted junction model into a classical nonlinear model\cite{wu2000NLZ,liu2002NLZ,liu2003NLZ}. We define $s_z=|\psi_1|^2-|\psi_0|^2$ and $\phi = \arg \psi_0 - \arg \psi_1$. The equations Eq.~(\ref{eq:linearSE}) and~(\ref{eq:RSJ}) are transformed to three classical equations,
\begin{eqnarray}
\dot \theta &&= \frac{1}{\tau_\theta} (1-I_1 \sin \theta - I_2 s_z \sin \theta/2)
 \\
 \dot \phi &&= \frac{1}{\tau_\phi} \cos \frac{\theta}{2} + \frac{s_z}{\tau_s\sqrt {1-s^2_z}}\cos \phi,
 \\
\dot s_z  &&= \frac{1}{\tau_s}\sqrt{1-s^2_z} \sin \phi, \label{eq:szdynamics}
\end{eqnarray}
where $I_1 = I_{\rm J}/I_{\rm in}$ and $I_2 = I_{\rm M}/I_{\rm in}$. We define three typical time scales $\tau_\theta = \hbar/2eRI_{\rm in}$, $\tau_\phi = \hbar / E_{\rm M}$, and $\tau_s = \hbar / \delta$. They determine the scale of velocity for the dynamics of the $\theta$, $\phi$, and $s_z$ respectively.

This mapping to a pure classical model enables analysis for the dynamics of $s_z$ through the method of averaging. Since $\delta$ is exponentially suppressed by the width of the junction and usually very small in the realistic junctions, we can treat $s_z$ as the slow variable and fix its value to solve Eqs. (4) and (5) first.
After obtaining the solution $\phi(t,s_z)$, we can then average $\sin \phi$ over a time period of $T$ with $\tau_\phi \ll T \ll \tau_s$, which gives a quantity $f(s_z) =\frac{1}{T} \int_{0}^{T} dt \sin \phi(t,s_z)$. Plugging it back to the Eq.~(\ref{eq:szdynamics}), we arrive at a self-consistent equation for the slow variable $s_z$ as
\begin{eqnarray}
\dot s_z = \frac{f(s_z)}{\tau_s}\sqrt{1-s^2_z}.
\end{eqnarray}
This equation determines the phase-space flow for $s_z$.

Two typical examples of such flow for the high and low voltage regime are shown in Fig.~\ref{fig:bifurcation}a, where the direction of the phase-space flow is indicated by small arrows. By obtaining the fix points with $\dot s_z=0$, we notice a fixed point at $s_z = 0$ for both cases. For the large voltage scenario, this fixed point is the only stable fixed point, which dominates the whole phase space. Any initial state for $s_z$ inevitably flows to $s_z=0$. The corresponding time evolution of $s_z$ is shown as the red curve of fig.~\ref{fig:bifurcation}b, which can be analytically described with a damped oscillating evolution as $s_z(t) \approx e^{-t/\tau_{\rm d}} \cos (t/ \tau_s)$ with $\tau_d \approx \tau_s \tau_\phi/\tau_\theta$ \cite{feng2018hysteresis}.
For the low voltage case, however, this fixed point at $s_z=0$ becomes an unstable one, which can be easily read out from the reversal of the flow directions in Fig.~\ref{fig:bifurcation}a. For this case, the system flows to the stable fixed point at $s_z \neq 0$, shown as the blue curve of Fig.~\ref{fig:bifurcation}, then the non-vanishing $f_{\rm J}/2$ radiation is expected.

To see the behavior of the phase-space flow more clearly, we demonstrate the portrait of the fixed points as a function of the excess voltage $V_{\rm ex} \equiv (I_{\rm in}-I_{\rm c})R$ in Fig.~\ref{fig:bifurcation}c, with $I_c$ the critical current above which the voltage emerges. 
Let us read this figure from the high-voltage to low-voltage regime. We first find that $s_z =0$ is the only stable fixed point at large voltages (red dots).
Below a critical excess voltage around $V_{\rm ex}=0.5E_{\rm M}/2e$, $s_z=0$ changes to an unstable fixed point. The system should flow to other stable fixed points at $s_z \neq 0$, which leads to non-vanishing $f_{\rm J}/2$ radiation. We notice that the fixed point at $s_z = 0$ switches its stability character multiple times at low voltages, which explains the interrupted line in Fig.~\ref{fig:setup}b. Another important feature of this fixed-point portrait is the generation of more fixed points when decrease the voltage. For clear demonstration, we trace the fixed points with lines and plot them as a function of $1/V_{\rm ex}$ in Fig.~\ref{fig:bifurcation}d. We find that the fixed points are generated with a process of splitting one fixed point into three fixed points. This process is known as the bifurcation. With decreasing voltage, the bifurcation generates more and more fixed points. When the fixed points become condensed enough that the system could freely wander around the stable ones and the chaotic dynamics naturally emerges.
This bifurcation is a standard route towards chaos, and qualitatively explains the noise-like signal at the bottom of the Fig.~\ref{fig:setup}b and Fig.~\ref{fig:spectra}.

\section{Conclusion}
In summary, we use a quantum resistively shunted junction model to study the Josephson radiation of a topological junction. We find that the $4\pi$-periodic Josephson radiation vanishes above a critical voltage, which explains a recent experiment in HgTe-based topological Josephson junction. We further predict additional interrupted emission lines and chaotic features in the radiation spectra, which provide a guidance for checking the structure of radiation spectra in topological Josephson junctions, and expect their verification by future experiments with a broader parameter range. 

{\it Acknowledgments.---}
The authors are grateful for Efstathios G. Charalampidis, Benedikt Fauseweh, James Williams, Zhong-Bo Yan and Jianxin Zhu for helpful discussions. This work was supported by NKRDPC-2017YFA0206203, 2017YFA0303302, 2018YFA0305603, and
NSFC (Grant No. 11774435). Zhao Huang is supported by U.S. DOE Office of Basic Energy Sciences E3B5. Qian Niu is supported by DOE (DE-FG03-02ER45958, Division of Materials Science and Engineering), NSF (EFMA-1641101) and Robert A. Welch Foundation (F-1255).

\appendix

\section{Model for Topological Josephson junctions based on quantum-spin-Hall insulator}
\begin{figure}[h]
\begin{center}
\includegraphics[clip = true, width =1\columnwidth]{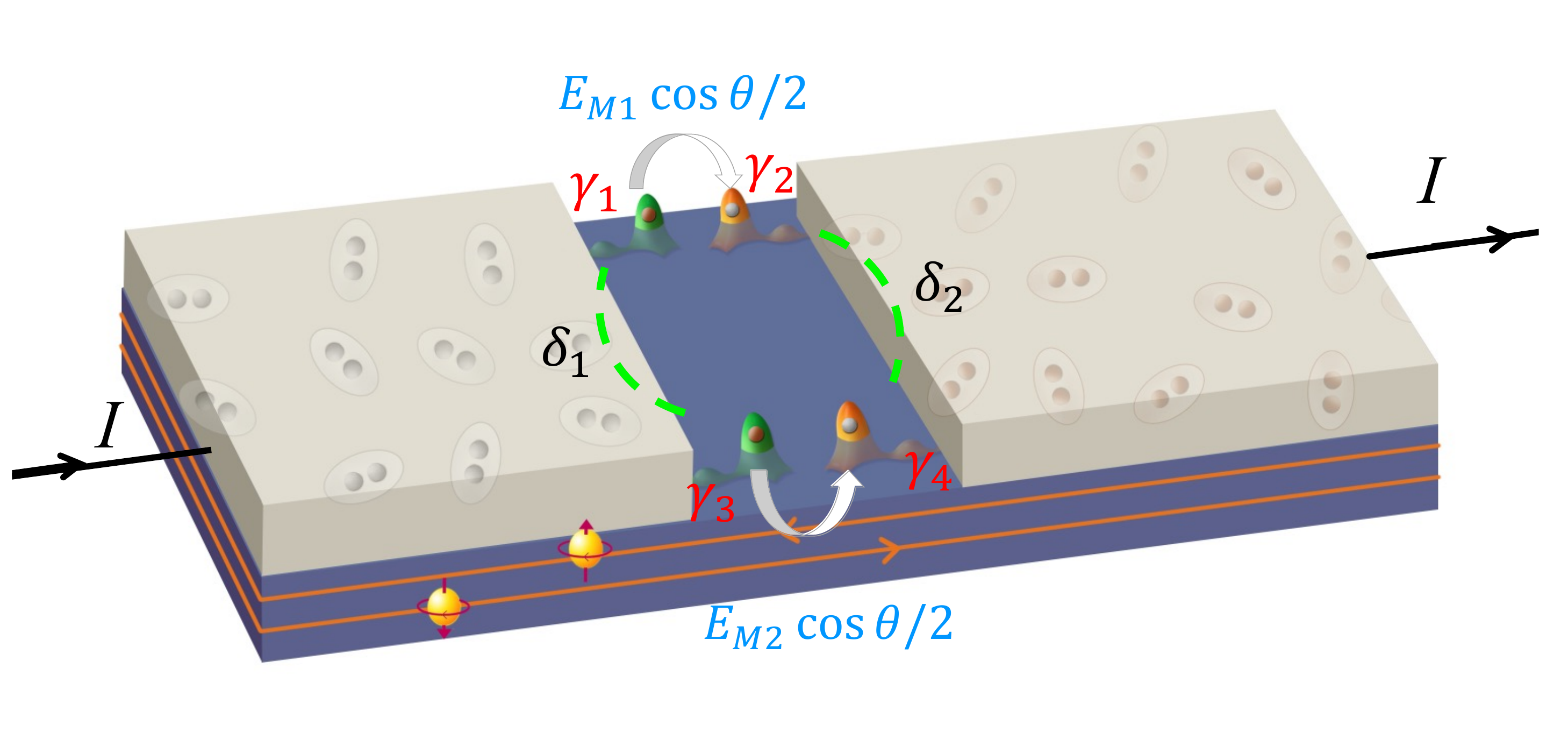}
\caption{ (a) Schematic setup of quantum-spin-Hall-insulator based topological Josephson junction. Four Majorana zero modes $\gamma_{1,2,3,4}$ exist at the junction, whose coupling contributes to the $4\pi$-periodic Josephson current.}
\label{fig:supplesetup}
\end{center}
\end{figure}

Topological Josephson junctions can be formed with different types of topological superconductors\cite{kitaev2001unpaired,beenakker2013,aguado2017,Sato2017,lutchyn2018review,zhou2019}. Our model for the topological Josephson junction is based on the experiment reported in Ref. [\onlinecite{deacon2017radiation}], where two superconductors are placed on a quantum-spin-Hall insulator as shown in Fig.~\ref{fig:supplesetup}. The superconductivity can penetrate into the insulator with the proximity effect and give rise to topological superconductivity. The supercurrent is thus transported through two distinct channels in presence of Majorana zero modes (MZMs). One is the conventional Cooper-pair channel with current phase relation $I_C= I_{\rm J} \sin \theta$, with $\theta$ the phase difference between the two superconductors and $I_J$ the maximum supercurrent\cite{mahan2000}. The other is the single-electron-tunneling channel mediated by the coupling between MZMs across the junction \cite{fu2009QSHIjunction}. Since there are two edges of the insulator attaching with the superconductors, there are two pairs of MZMs as shown in Fig.~1: $\gamma_1$ and $\gamma_2$ at the upper edge, and $\gamma_3$ and $\gamma_4$ at the lower edge. The two edges together contribute the Josephson current 
\begin{eqnarray}\label{eq:4picurrent}
I = \left(I_{\rm M1} \langle i\gamma_1 \gamma_2 \rangle+ I_{\rm M2}\langle i\gamma_3 \gamma_4\rangle \right) \sin \theta/2.
\end{eqnarray} 

In the following we calculate the Josephson current through the Majorana channel with the Hamiltonian
\begin{eqnarray}
H_M=H_J+H_\delta
\end{eqnarray}
where 
\begin{eqnarray}\label{eq:Josephson}
H_J = - E_{\rm M1}  i\gamma_1 \gamma_2 \cos (\theta/2) - E_{\rm M2}  i\gamma_3 \gamma_4 \cos (\theta/2),
\end{eqnarray}
is the Hamiltonian of the Josephson coupling across the upper and lower edge, and
\begin{eqnarray}\label{eq:delta}
H_\delta = - \delta_1 i\gamma_1 \gamma_3 - \delta_2 i\gamma_2 \gamma_4,
\end{eqnarray}
is the Hamiltonian for the small couplings between $\gamma_1$ and $\gamma_3$, the two MZMs on the left side of the barrier, and between $\gamma_{2}$ and $\gamma_4$, the two MZMs on the right side of the barrier. The coupling coefficient $\delta_1$ and $\delta_2$ are small but nonzero because the overlap between the wavefunction of MZMs exponentially decays with their distances. In realistic junctions, $\delta_1$ and $\delta_2$ are much smaller than other energy scales of the dynamics in this junction such as $E_{M1}$, $E_{M2}$ and injected energy from the external current. 

Because two MZMs correspond to one electron, we can define the electron operators $c_1=(\gamma_1+i\gamma_2)/2$ and $c_2=(\gamma_4+i\gamma_3)/2$, which lead to 
\begin{equation}
\begin{aligned}
&i\gamma_1\gamma_2=2c_1^\dagger c_1 -1, \\
&i\gamma_3\gamma_4=1-2c_2^\dagger c_2, \\
&i\gamma_1\gamma_3=c_1c_2-c_1c_2^\dagger+c_1^\dagger c_2-c_1^\dagger c_2^\dagger, \\
&i\gamma_2\gamma_4=c_1c_2+c_1c_2^\dagger-c_1^\dagger c_2-c_1^\dagger c_2^\dagger.
\end{aligned}
\end{equation}
By putting the above results into Eq.~(\ref{eq:Josephson}) and (\ref{eq:delta}), in the Fock space with the basis $|00\rangle$, $|11\rangle$, $|01\rangle$ and $|10\rangle$, the Hamiltonian matrix is given by
\begin{small}
\begin{eqnarray}
&&\mathcal{H}=
\\\nonumber
&&\left(\begin{smallmatrix}
(E_{{\rm M1}} - E_{{\rm M2}})\cos(\theta/2) & \delta_{1}+\delta_{2} & 0 & 0\\
\delta_{1}+\delta_{2} & (E_{{\rm M2}} - E_{{\rm M1}})\cos(\theta/2) & 0 & 0\\
0 & 0 & (E_{{\rm M1}} + E_{{\rm M2}})\cos(\theta/2) & -\delta_{1}+\delta_{2}\\
0 & 0 & -\delta_{1}+\delta_{2} & -(E_{{\rm M1}} + E_{{\rm M2}})\cos(\theta/2)
\end{smallmatrix}\right),
\end{eqnarray}
\end{small}
where the up left and lower right block corresponds to even and odd fermionic parity respectively.  The elements of off-diagonal block are all zero, which reflects conservation of the total parity in this system. Without losing generality, we take the odd total parity, and the Hamiltonian is thus simplified to
\begin{equation}
\mathcal{H}=\left(\begin{matrix}
E_{{\rm M}}\cos(\theta/2) & \delta\\
\delta & -E_{{\rm M}}\cos(\theta/2)\\
\end{matrix}\right)
\end{equation}
with $E_{{\rm M}} = E_{{\rm M1}} + E_{{\rm M2}}$ and $\delta = \delta_{2}-\delta_{1}$. This is the Hamiltonian shown in Eq.~(2) of the main text. The general Majorana wave function can be written as $|\psi \rangle = \psi_0 |01\rangle + \psi_1 |10\rangle$, and thus the current in Eq.~(\ref{eq:4picurrent}) is given by
\begin{eqnarray}
I = I_{\rm M} (|\psi_1|^2 - |\psi_0|^2) \sin \theta/2,
\end{eqnarray}
with $I_{\rm M} = I_{\rm M1} + I_{\rm M2}$. This is the $4\pi$-periodic Josephson current in Eq.~(3) of the main text.

\section{Proof of quantized Josephson radiation frequencies}
Now we prove that the Josephson radiation has quantized frequencies $f= nf_J/2$ with a dc bias current source when $|\psi_1|^2 - |\psi_0|^2 = s_{z0}$, a fixed nonzero value. In the resistively shunted junction (RSJ) model of this topological Josephson junction, the external current $I_{in}$ is divided into three parts: the dissipation current, tunneling current through the Cooper-pair channel and the Majorana channel. After considering the ac Josephson effect $V(t) = \hbar \dot \theta(t)/ 2e$, we obtain the equation for the Ohm's law as
\begin{eqnarray}\label{eq:RSJ}
\dot \theta = \omega_J (1-I_1 \sin \theta - I_2 \sin \theta/2),
\end{eqnarray}
where $\omega_J = 2eRI_{in}/\hbar$, $I_1 = I_{\rm J}/I_{in}$ and $I_2 = I_{\rm m} s_{z0}/I_{in}$ are fixed parameters \cite{tinkham2004}. Since the external current is fixed, the radiation frequencies are fully determined by the Fourier components of the time-dependent voltage.

Let us first demonstrate that $V(t)$ is a periodic function of time. The Eq.~(\ref{eq:RSJ}) shows that $V$ is a $4\pi$-periodic function of the Josephson phase $V(\theta) = V(\theta + 4\pi)$. Now let us examine the time $T$ for $\theta$ to increase $4\pi$ from an arbitrary initial phase $\theta_0$. We have
\begin{eqnarray}
T &&= \int_{\theta_0}^{\theta_0+4\pi} d\theta /\dot \theta = \int_{\theta_0}^{\theta_0+4\pi} \frac{d\theta} {\omega_J (1-I_1 \sin \theta - I_2 \sin \theta/2)}
\nonumber\\
&&=\left(\int_{\theta_0}^{4\pi}  + \int_{4\pi}^{\theta_0+4\pi}\right)\frac{d\theta} {\omega_J (1-I_1 \sin \theta - I_2 \sin \theta/2)} 
 \\\nonumber
&&=\left(\int_{\theta_0}^{4\pi}  + \int_{0}^{\theta_0}\right)\frac{d\theta} {\omega_J (1-I_1 \sin \theta - I_2 \sin \theta/2)} 
 \\
&&=\int_{0}^{4\pi} \frac{d\theta} {\omega_J (1-I_1 \sin \theta - I_2 \sin \theta/2)},
\end{eqnarray}
which shows that $T$ is independent of the initial phase $\theta_0$. Therefore we obtain the time periodicity of the voltage $V(t) = V(t+T)$. 

Now we show that $T = 2/f_J$ with $f_J=2eV_0/\hbar$. Here $V_0$ is the average voltage. We have
\begin{equation}
f_J = 2eV_0/h = \frac{2e}{ h T} \int_0^T  dt V(t)
= \frac{1}{2\pi T} \int^{4\pi}_0 d \theta = 2 /T,
\end{equation}
which indicates that the voltage has the time period $2/f_J$. Based on this time periodicity, we obtain the Fourier series as
\begin{equation}
V(t) = V_0 + \sum _n v_n e^{i \pi n f_J  t},
\end{equation}
which demonstrates the quantized frequencies $f = nf_J/2$ of the Josephson radiation.

\begin{figure}[h]
\begin{center}
\includegraphics[clip = true, width =1\columnwidth]{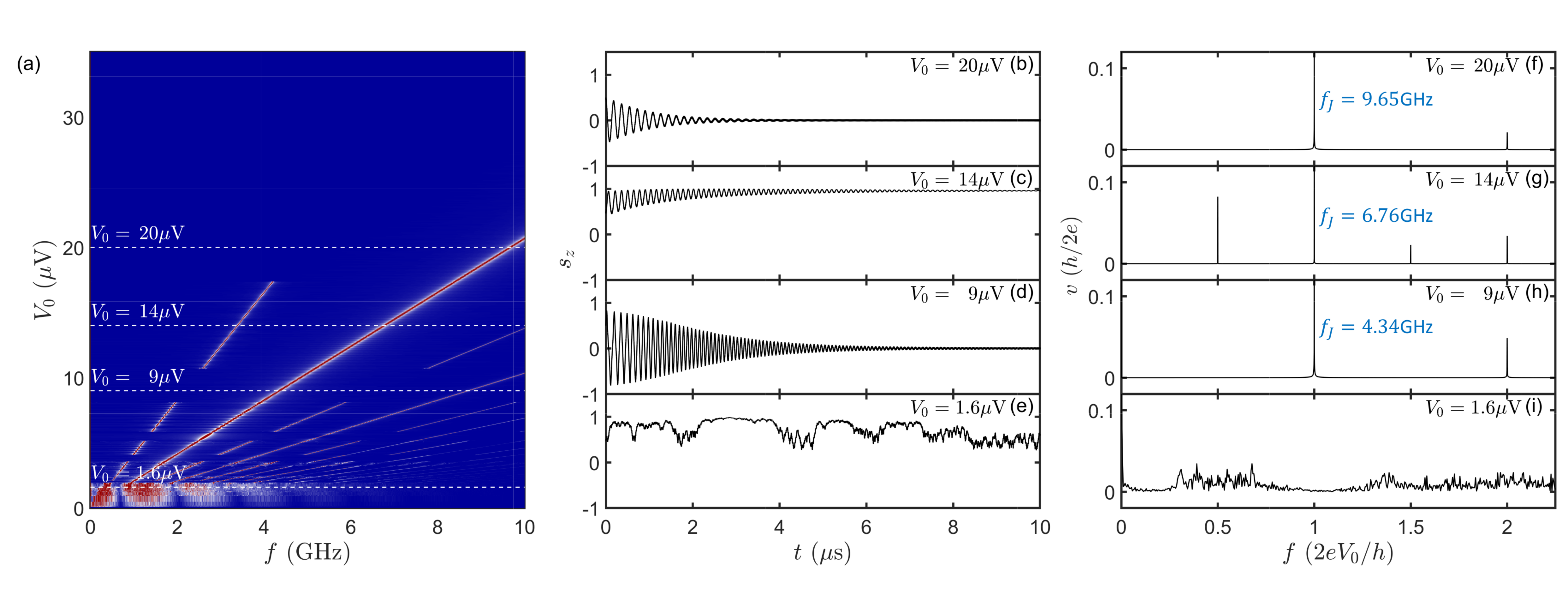}
\caption{ (a) Numerical results of the Josephson radiation with parameters the same as in Fig.~1b of the main text, where the dashed lines, from top to bottom, indicate the typical voltages of $V_0 = 20\mu$V, $14\mu$V, $9\mu$V, and $1.6\mu$V, respectively. (b-e) The time evolution of the $s_z$ for the four typical voltages. (f-i) The radiation power spectra for the typical voltages. }
\label{fig:numerical}
\end{center}
\end{figure}

\section{The quantum dynamics of the two-level system formed by Majorana zero modes}
Here we present numerical details on the simulations of quantum resistively shunted junction model, which lead to Fig.~\ref{fig:setup}b of the main text. The model contains two coupled equations of,
\begin{eqnarray}\label{eq:linearSE}
&&i \hbar \frac{{\rm d}}{{\rm d} t} \begin{pmatrix}
\psi_0  \\
\psi_1
\end{pmatrix} = \begin{pmatrix}
E_{\rm M}\cos{\frac{\theta}{2} } & \delta\\
\delta & -E_{\rm M}\cos{\frac{\theta}{2} }
\end{pmatrix} \begin{pmatrix}
\psi_0  \\
\psi_1
\end{pmatrix}
\nonumber\\
&&\dot \theta = \frac{2eR}{\hbar} \left[I _{\rm in}- I_{\rm J} \sin \theta - I_{\rm M} s_z \sin \frac {\theta}{2}\right]
\end{eqnarray}
with $s_z=|\psi_0|^2-|\psi_1|^2$, which are Eqs (2) and (3) of the main text. Within these two coupled equations, $\psi_{0,1} (t)$ and $\theta (t)$ are the solutions to be obtained through simulation, $(R, I_{\rm M}, I_{\rm J}, E_{\rm M}, \delta)$ are fixed parameters determined by the construction of the junction. The only controlled variable is the input current $I_{\rm in}$. To obtain the radiation spectrum, we consider a realistic scenario, in which the input current $I_{\rm in}$ is adiabatically increased from zero to a large value. For each $I_{\rm in}$, we use the Runge-Kutta method to simulate the coupled equations to obtain the Josephson phase $\theta (t)$ and the voltage $V(t,I_{\rm in}) = \hbar \dot \theta/2e$. As shown in Fig.~\ref{fig:numerical}(b) to (d), the $s_z$ can converge to a specific value. After the convergence, we use the Fourier transformation to obtain $v (f,I_{\rm in})=\int_0^T V(t,I_{\rm in})e^{i 2\pi ft}$. The dc component is given by $V_0 = v(0,I_{\rm in})$, which is a monotonic function and has a one-to-one correspondence between the dc voltage $V_0$ and the input current $I_{\rm in}$. By scanning $f$, we can obtain $v(f,V_0)$, which is drawn in Fig.~\ref{fig:numerical}a. The shinny lines indicate the quantized frequencies where the intensity $v(f,V_0)$ has a peak. 

A set of realistic junction parameters $(R, I_{\rm M}, I_{\rm J}, E_{\rm M}, \delta)$ provides the results shown in Fig.~1b of the main text. For convenience we show the numerical result again in Fig.~\ref{fig:numerical}a. We draw four white dashed lines to indicate four typical dc voltages $V_0 = 20\mu$V, 14$\mu$V, 9$\mu$V, 1.6$\mu$V. For $V_0 = 20\mu, 9 \mu V$, the wave function average $s_z$ experiences a damped oscillation towards zero, as shown in Fig.~\ref{fig:numerical}b and \ref{fig:numerical}d. For these cases, there is no $f_J/2$ radiation due to vanishment of tunneling current through the Majorana channel, as shown in Fig.~\ref{fig:numerical}f and \ref{fig:numerical}h. However for $V_0 = 14\mu$V in Fig.~\ref{fig:numerical}c, $s_z$ converges to a nonzero value. The tunneling current through the Majorana channel exist and induce the Josephson radiation with frequencies $nf_J/2$ as shown in Fig.~\ref{fig:numerical}g.
Finally, we show the results for the low voltage limit with $V_0 = 1.6\mu$V. From Fig.~\ref{fig:numerical}e, we see that the dynamics of $s_z$ is chaotic, which gives a noise-like spectra as shown in Fig.~\ref{fig:numerical}i, with no clear peaks at quantized frequencies.

\bibliography{Feynman}

\end{document}